\documentclass[journal,12pt,onecolumn,draftclsnofoot,]{IEEEtran}
\usepackage{graphicx}
\usepackage{subcaption}
\usepackage{cite}
\usepackage{multirow}
\usepackage{amssymb}
\usepackage{scalerel}
\usepackage{algorithm}
\usepackage{algpseudocode}
\normalsize

\usepackage{amsthm}

\usepackage[cmex10]{amsmath}
\usepackage [english]{babel}
\usepackage [autostyle, english = american]{csquotes}
\MakeOuterQuote{"}

\begin{document}
\title{Energy Efficiency Optimization for Device-to-Device Communication Underlaying Cellular Networks in Millimeter-Wave}

\author{Negar~Zabetian,
	Abbas~Mohammadi,
	and~Mohammad~Kazemi}

\maketitle
\begin{abstract}
This paper studies energy efficiency maximization in device-to-device (D2D) communications underlaying cellular networks in millimeter-wave (mm-wave) band. A stochastic geometry framework has been used to extract the results. First, cellular and D2D users are modeled by independent homogeneous Poisson point process; then, exact expressions for successful transmission probability of D2D and cellular users have been derived. Furthermore, the  average sum rate and energy efficiency for a typical D2D scenario have been presented. An optimization problem subject to transmission power and quality of service constraints for both cellular and D2D users has been defined and energy efficiency of D2D communication is maximized. 
Simulation results reveal that by working in millimeter-wave, significant energy efficiency improvement can be attained, e.g., 20\% energy efficiency improvement compared to Rayleigh distribution in the practical scenarios by considering circuit power. Finally, to verify our analytical expressions, the simulation studies are carried out and the excellent agreements have been achieved.	
\end{abstract}

\IEEEpeerreviewmaketitle
\section{Introduction}

\IEEEPARstart It is predicted that mobile data becomes 1000 fold by 2020 \cite{1}. To achieve this objective, 3GPP has standardized new technologies with a high data rate, low latency, and less power consumption. Accordingly, a new generation of mobile communications known as fifth generation (5G) is proposed. Compared to 4G, 5G is expected to have higher capacity and throughput and lower latency. New technologies, such as massive multiple input multiple output (MIMO), non-orthogonal multiple access (NOMA), spatial modulation and device-to-device (D2D) communications have been considered to achieve 5G requirements \cite{2}. Moreover, D2D communication in which devices with short distances connect  directly without any infrastructure, imposing less traffic on the core of the network. This will reduce energy consumption. D2D communications, due to the proximity of devices, have some advantages such as high spectral efficiency (SE), high energy efficiency (EE), low transmission power, high bit rate, and low latency \cite{3}.

D2D communications are classified into inband (licensed) and outband (unlicensed). In the licensed communications which are divided into underlay and overlay modes, devices use cellular spectrum. In  underlay communications, the same channels that are allocated to cellular users, are used by devices. Thus D2D users interfere with cellular users. In an overlay mode, some channels from the cellular spectrum are dedicated to devices, so they do not suffer from co-channel interference. Nevertheless, spectral efficiency is not as efficient as underlay mode \cite{4}.

Since in an underlay scheme devices use the same spectrum as cellular users, these communications face new challenges. Thus the interference between the cellular and D2D users should be managed, and power allocation becomes a significant problem. There are several methods to solve resource allocation problems such as game theory, graph theory, and stochastic geometry. Game theory is used as a mathematical tool for modeling D2D communications in \cite{5}. Graph theory for interference management in D2D communications underlaying cellular networks is used in \cite{6}. Stochastic geometry is a powerful tool for modeling wireless networks which uses point process theory \cite{7}. The stochastic optimization problem for D2D power allocation is formulated in \cite{8} which leads to computing D2D ergodic rate. There are several works which use Poisson point processes (PPPs) to model cellular and D2D users \cite{9}, \cite{10}.

Furthermore, green communications have attracted a lot of attention recently. Energy storage through the green network reduces CO$_2$ emissions and thus reduces global warming. There are also incentives to reduce energy consumption in wireless networks. EE is a performance metric in green communication \cite{11}. Extensive researches have been done in EE maximization in D2D communication \cite{12}, \cite{13}. EE and SE trade-off is studied in \cite{14}. The authors in \cite{15} study EE maximization in D2D communications on multiple bands, propose derivative based algorithms and use stochastic geometry approach. Moreover, a closed-form expression for spectral efficiency is obtained in \cite{16} and EE is maximized in cellular networks by deploying stochastic geometry. 

On the other hand, millimeter-wave (mm-wave) is another proposed key technology in 5G. Its operation frequency varies from 30GHz to 300GHz. So this large bandwidth becomes attractive for cellular networks. By using mm-wave, the antenna size becomes smaller, and it is possible to pack multiple antenna elements in a small area at transmitters and receivers. One of the most important problem in mm-wave is the blockage,  in which, non-line of sight (NLOS) paths become weaker, and they cannot penetrate objects well. However, the directional beamforming at transmitter and receiver allows high quality links \cite{17}. The performances of the line of sight (LOS) and NLOS transmission in a cellular network with Rician fading has been studied in \cite{18}. Cellular network at mm-wave frequency is modeled in \cite{19} and \cite{20}. Also,  coverage and rate have been analyzed by deploying stochastic geometry in these studies. 

Utilizing mm-wave in D2D communication because of its short range is practical.  Combination of D2D communication and mm-wave improves the performance of wireless network \cite{sim20175g}. Performance of mm-wave D2D networks using the Poisson bipolar model is investigated in  \cite{21}.  The ergodic rate of Ad-Hoc networks at the mm-wave range is investigated in \cite{23}.   An analytical framework to analyze the uplink performance of D2D communication in mm-wave  network is provided in \cite{turgut2017uplink} by using tools from stochastic geometry. A flexible mode selection scheme and Nakagami fading is employed to analyze outage probability. The spectral and  energy efficiency of outband D2D users with directional mm-wave antennas are investigated in \cite{chevillon2018spectral}, where the transmission power of D2D users is relative to the D2D link distance.  In \cite{trigui2018unified}, the cellular and D2D SINR distributions are evaluated in general fading conditions e.g. Nakagami-$m$ by using stochastic geometry approach. The average area spectral efficiency utility of D2D communication is maximized while coverage probability of cellular users is guaranteed.

In this paper, we consider a multiple-band uplink cellular network with D2D and cellular users at the mm-wave frequency. Furthermore,  the challenges of the mm-wave channel such as beamforming directivity, blockage, and high attenuation are investigated. We model the locations of D2D and cellular users as independent homogeneous PPPs. Then, we study energy efficiency as a performance metric in green communication using stochastic geometry approach. 

More specifically, our contributions are as follows:
\begin{itemize}
	\item 	An analytical framework to study the uplink performance of underlay D2D communication by using stochastic geometry tools is introduced  in millimeter-wave network.
	\item  Considering blockage and directional beamforming at transmitter and receiver, Laplace transform expressions for both cellular and D2D interference links are obtained.  Then,	signal to interference plus noise ratio (SINR) expressions for both cellular and D2D users are provided and successful transmission probability for these users are computed. 
	\item Closed-form formulas for average sum rate (ASR) and EE of D2D users are obtained.  Then, we formulate an optimization problem to maximize the energy efficiency of D2D users while considering the QoS of both cellular and D2D users. Finally, the optimal transmission power of D2D users is obtained.  
\end{itemize}
The rest of this paper is organized as follows. The system model is presented in Section II. In Section III, we derive analytical formulas for ASR and EE of D2D users and define the main optimization problem  to obtain the optimal transmission powers for D2D users. Simulation and numerical results are presented in Section IV. Finally, Section V represents conclusions and some future directions. 

Notation: $\mathbb{P}(.)$, $\mathbb{E}(x)$, $\rm\Gamma(.)$  and $\exp (.)$ denotes the probability, expectation value of a random variable $x$, gamma distribution and exponential function, respectively.

\section{System model} \label{sec2}
\label{sec:1}
We consider D2D communications underlaying an uplink cellular network in which devices can reuse the same spectrum which is used by cellular users. In our network, we have multiple bands by dividing the entire spectrum to M bands. Each band is denoted by subscript $i$ ($i=1, 2,\dots, M$) and $W_{i}$ represents the bandwidth of $ i$-th band. In each band, the number of cellular users and devices are modeled by independent homogeneous PPPs. Also, we consider that signals in different bands do not interfere with each other. As can be seen in Fig. 1, all the devices and cellular users are on a two-dimensional plane $\mathbb{R}^2$. To differentiate the cellular users from D2D users, superscripts C and D have been used, respectively throughout the paper.

\begin{figure}[ht]
	\centering
	\includegraphics[width=.5\linewidth]{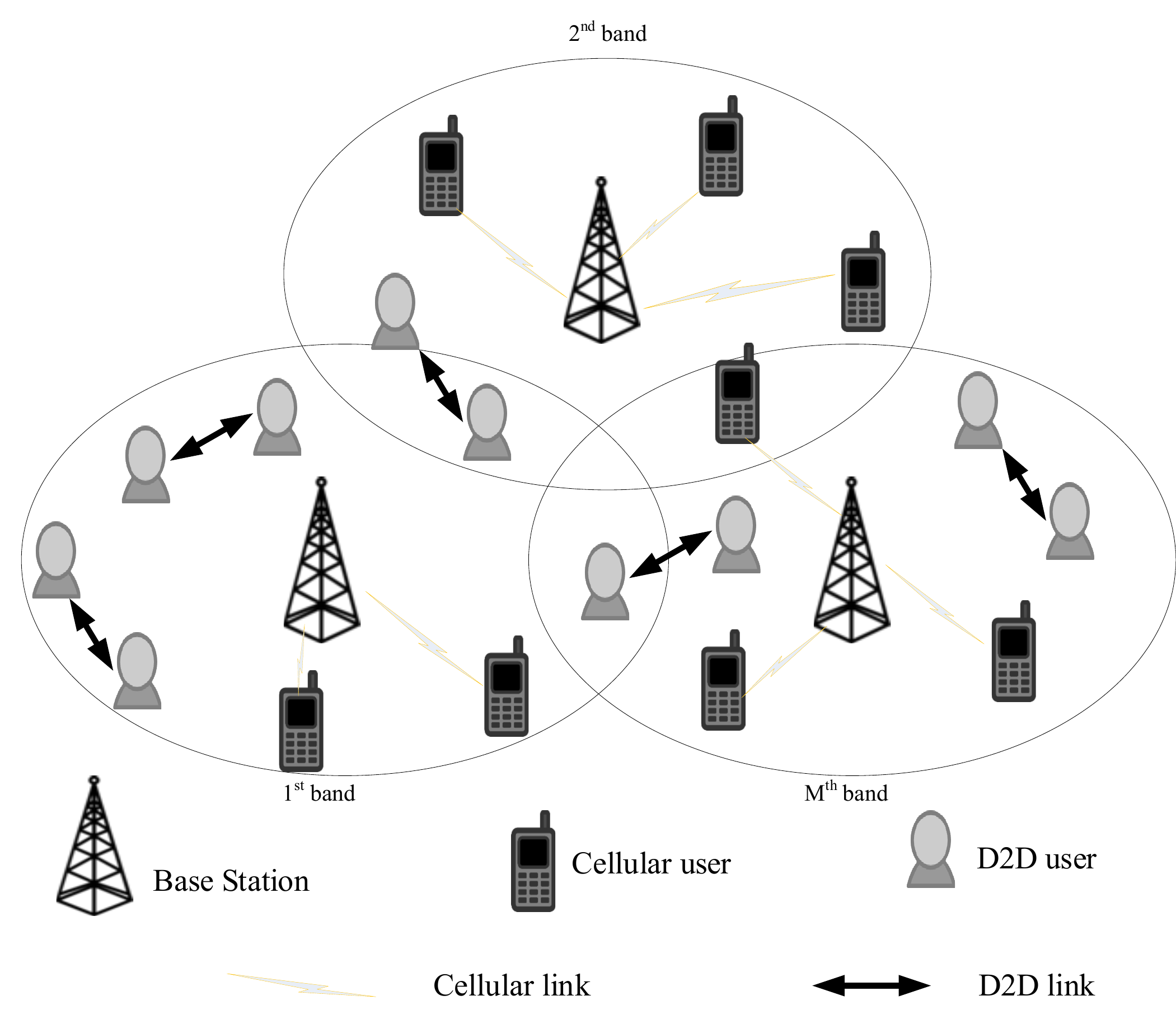}
	\caption{An illustration of D2D network underlaying cellular communication.}
	\label{fig.1}
\end{figure}

According to Palm theory \cite{24}, for Poisson point process, we have a typical base station (BS) for cellular communication and a typical D2D receiver for D2D communication at origin on $\mathbb{R}^2$. It means that all points have equal chance to be chosen as a typical user. Also, conditioning on a point does not have an impact on the distribution of rest of them, and they still have PPP distribution due to Slivnyak's theorem \cite{24}.

Directional beamforming is assumed in our system model. Antenna arrays at the transmitters and receivers are considered to perform directional beamforming. It means that the main lobe is directed towards the dominant path and side lobes lead energy in other directions. The array patterns are approximated by a sectored antenna model \cite{25}. Perfect beam alignment is considered between the transmitters and the receivers. So, an overall antenna gain in perfect alignment is equal to $GG$ where $G$ is the main lobe gain both at the transmitter and the receiver sides. Also, the beam direction of the interfering users is modeled by uniform distribution on $[0,\,2\pi )$. Thus, the effective antenna gain is a discrete random variable with the probability distribution described by  \cite{thornburg2016performance}
\begin{equation}\label{eq:1}
{G_k} = \left\{ \begin{array}{l}
GG,\,\,\,\,\,w.p.\,\,\,\,p_{GG} = {\big( {\frac{\theta }{{2\pi }}} \big)^2}\\
Gg_s,\,\,\,\,\,\,w.p.\,\,\,\,p_{Gg_s} = 2\big( {\frac{\theta }{{2\pi }}} \big)\,\big( {\frac{{2\pi  - \theta }}{{2\pi }}} \big)\,\,\,\,\,\,\,\,\,\\
g_sg_s,\,\,\,\,\,\,w.p.\,\,\,\,p_{g_sg_s} = {\big( {\frac{{2\pi  - \theta }}{{2\pi }}} \big)^2}\,\,\,
\end{array} \right.
\end{equation}  
where $g_s$ is the side lobe gain, $\theta $ is the beam width of the main lobe and $p_{G_k}$ is the probability of having a combined antenna gain of ${G_k}$.

For analytical tractability, the mm-wave small-scale fading is modeled by Nakagami distribution with parameter $m$ which is more general than Rayleigh fading \cite{19}   and all of the mm-wave equations are in terms of Nakagami parameter. Therefore, channel gain has a gamma distribution, which is denoted by  $g\sim\rm\Gamma (m,\frac{1}{m})$. By considering, small- and large-scale fading, the received power can be written as 
\begin{equation}\label{eq:2}
{P_r} = {P_t}GGg{r^{ - \alpha }}, 
\end{equation}
where $P_{t}$  is the transmission power. Also, $ {r^{ - \alpha }} $ represents the large-scale path loss in which $r$ is the distance between transmitter and receiver, and $\alpha$ is path loss exponent. The signal path can be line-of-sight (LOS) or non-light-of-sight (NLOS). If there are no blockages between the transmitter and receiver link, the path is LOS. Otherwise, it is NLOS. Mm-wave propagation measurements in \cite{26}, show that path loss exponent in LOS and NLOS is different.  So, we
consider different path loss exponents for LOS and NLOS. The probability of the LOS link with length of $r$ is  given by ${f_L}\big( r \big)=\exp ( - \beta r)$ where $\beta$ depends on the building parameters and density \cite{27}. Also, the probability of the NLOS link is ${f_N}\big( r \big) = 1 - {f_L}\big( r \big)$. To differentiate the LOS from NLOS, superscripts L and N have been used, respectively throughout the paper.

Sum of cellular user's powers is constant, and these users in each band have the same power $P_{c,i}$. So we have $P_{c,i}=P_{c}/M$
where $P_{c}$ is the total transmission power of cellular users, similarly, the devices have total transmission power $P_{d}$ and devices in $ i$-th band have same power level $P_{d,i}$.
\begin{equation}\label{eq:3}
\sum\limits_{i = 1}^M {{P_{d,i}} \le {P_d}}.
\end{equation}
Moreover, the power of the D2D transmitter in the $i$-th band is bounded as follows. 
\begin{equation}\label{eq:4}
0 \le {P_{d,i}} \le {P_{d,i,\max }}.
\end{equation} 
where ${P_{d,i,\max }}$ is a specific threshold for the power of D2D transmitters in each band.

The received message at a typical D2D user in $ i$-th band is 
\begin{equation}\label{eq:5}
\begin{array}{l}
{Y_{d,i}}= {\!X_{d,i}}{h_{d,00}}\sqrt {{G_{d,00}}}R_{d,00,i}^{ - \frac{\alpha }{2}} + \sum\limits_{j \in {\phi_{c,i}}} {\!{X_{c,i}}{h_{c,j0}}\sqrt {{G_{c,j0}}}R_{c,j0,i}^{ - \frac{\alpha }{2}}} 
+ \sum\limits_{\ell  \in {\phi_{d,i}\backslash \{ 0\}}} {\!{X_{d,i}}{h_{d,\ell 0}}\sqrt {{G_{d,\ell 0}}}R_{d,\ell 0,i}^{ - \frac{\alpha }{2}}} \!+ n,  
\end{array}
\end{equation}
where $G_{d,00}$ is the  effective antenna gain in which main beams of typical D2D transmitter and receiver are aligned together, $G_{c,j0}$ and $ G_{d,\ell 0}$ are the  effective antenna gain of cellular and D2D interferes on typical D2D receiver, respectively, ${X_{d,i}}$ and ${X_{c,i}}$ are the information signals of D2D and cellular transmitters in the $ i$-th band, respectively and ${h_{d,00}}$ and ${R_{d,00,i}}$ are the channel fading coefficient and the distance between typical {D2D} transmitter and corresponding receiver in the $ i$-th band, respectively. Similarly, ${h_{d,\ell 0}}$ and ${R_{d,\ell 0,i}}$  state the channel fading coefficient and the distance between $ \ell$-th D2D transmitter and typical D2D receiver in the $ i$-th band, respectively, and ${h_{c,j 0}}$ and ${R_{c,j 0,i}}$  represent the channel fading coefficient and the distance between $ j$-th cellular transmitter and typical D2D receiver in the $ i$-th band, respectively. $ \phi _{c,i}$ and $ \phi _{d,i}$  denote the set of cellular users and devices in the $i$-th band, respectively and $n$ is additive white Gaussian noise (AWGN) with zero mean and variance $N_{0}$.

Therefore, by considering interference from D2D and cellular transmitters, the received SINR for typical D2D user in $ i$-th band is obtained.
\begin{equation}\label{eq:6}
\begin{array}{l}
SIN{R_{d,i}}
= \frac{{{P_{d,i}}{g_{d,00}}{G_{d,00}}R_{d,00,i}^{ - \alpha }}}{{\sum\limits_{j \in {\phi _{c,i}}} \!\!\! {P_{c,i}}{g_{c,j0}}{G_{c,j0}}R_{c,j0,i}^{ - \alpha } + \sum\limits_{\ell  \in {\phi _{d,i}\backslash \{ 0\}}}  \!\!\!{P_{d,i}}{g_{d,\ell 0}}{G_{d,\ell 0}}R_{d,\ell 0,i}^{ - \alpha } + {N_0}}}
\end{array},
\end{equation}         
where ${g_{d,00}}$ is the channel gain between a typical D2D pair, ${g_{c,j0}}$ is the channel gain between $ j$-th cellular transmitter and a typical D2D receiver and ${g_{d,\ell 0}}$ is the channel gain between $ \ell$-th D2D transmitter and typical D2D receiver in the $ i$-th band. We define the interference from cellular users to typical D2D receiver in $ i$-th band as ${I_{d,c0,i}}$. Also, ${I_{d,d0,i}}$  shows the interference from D2D transmitters to typical D2D receiver. So, $SIN{R_{d,i}}$ can be rewritten as
\begin{equation}\label{eq:7}
SIN{R_{d,i}} = \frac{{{P_{d,i}}{g_{d,00}}G_{d,00}R_{d,00,i}^{ - \alpha }}}{{{I_{d,c0,i}} + {I_{d,d0,i}} + {N_0}}}.
\end{equation} 
As the same way, the received signal by a typical BS is given by
\begin{equation}\label{eq:8}
\begin{array}{l}
{Y_{c,i}}
= {X_{c,i}}{h_{c,00}}\sqrt {{G_{c,00}}} R_{c,00,i}^{ - \frac{\alpha }{2}} + \sum\limits_{j \in {\phi _{c,i}}\backslash \{ 0\}} {{X_{c,i}}{h_{c,j0}}\sqrt {{G_{c,j0}}} R_{c,j0,i}^{ - \frac{\alpha }{2}}} + \sum\limits_{\ell  \in {\phi _{d,i}}} {{X_{d,i}}{h_{d,\ell 0}}\sqrt {{G_{d,\ell 0}}} R_{d,\ell 0,i}^{ - \frac{\alpha }{2}}}  + n.
\end{array}
\end{equation}
The SINR for typical BS in $ i$-th band can be written as follows.
\begin{equation}\label{eq:9}
\begin{array}{l}
SIN{R_{c,i}}= \frac{{{P_{c,i}}{g_{c,00}}{G_{c,00}}R_{c,00,i}^{ - \alpha }}}{{\sum\limits_{j \in {\phi _{c,i}}\backslash \{ 0\}} {{P_{c,i}}{g_{c,j0}}{G_{c,j0}}R_{c,j0,i}^{ - \alpha }}  + \sum\limits_{\ell  \in {\phi _{d,i}}} {{P_{d,i}}{g_{d,\ell 0}}{G_{d,\ell 0}}R_{d,\ell 0,i}^{ - \alpha }}  + {N_0}}},
\end{array}
\end{equation}
where $\sum\limits_{j \in {\phi_{c,i}}\backslash \{ 0\}} {{P_{c,i}}{g_{c,j0}}{G_{c,j0}}R_{c,j0,i}^{ - \alpha }} $  states the interference from cellular users to typical BS in $ i$-th band which is denoted by ${I_{c,c0,i}}$, $\sum\limits_{\ell  \in {\phi_{d,i}}} {{P_{d,i}}{g_{d,\ell 0}}{G_{d,\ell 0}}R_{d,\ell 0,i}^{ - \alpha }}$ expresses the interference from D2D transmitters to typical BS which is represented by ${I_{c,d0,i}}$ and ${g_{c,00}}$ is the channel gain between a typical BS and the cellular transmitter in $ i$-th band.
\section{Problem formulation} \label{sec3}
The performance metric of this network is energy efficiency (bit/J). EE is the ratio of average sum rate (ASR) to total power consumption \cite{28}. 

To calculate ASR, first, we should obtain the average rate from Shannon capacity. For this purpose, we use the lower bound on the rates of D2D and cellular users \cite{29}.
\begin{equation}\label{eq:10}
\begin{array}{l}
{{\bar R}_{d,i}} = \mathop {\sup }\limits_{{T_{d,i}} \ge 0} {W_i}lo{g_2}(1 + {T_{d,i}})\mathbb{P} (SIN{R_{d,i}} \ge {T_{d,i}}),\\
{{\bar R}_{c,i}} = \mathop {\sup }\limits_{{T_{c,i}} \ge 0} {W_i}lo{g_2}(1 + {T_{c,i}})\mathbb{P} (SIN{R_{c,i}} \ge {T_{c,i}}),
\end{array}
\end{equation}
where  ${T_{d,i}}$ and ${T_{c,i}}$ are SINR thresholds for D2D and cellular users, respectively. Also, to compute the average rate, the successful transmission probability (STP) for both cellular and D2D users should be obtained. For this purpose, first, the following Lemma is expressed. 

\textit{lemma 1.} \label{lemma1} If $y$ is normalized random variable with gamma distribution with parameter $m$, for a constant $z > 0$, the cumulative distribution function tightly bound as 
	\begin{equation}\label{eq:11}
	{(1 - {e^{ - az}})^m} < \mathbb{P} (y < z).
	\end{equation}
	with $a \buildrel \Delta \over = m{(m!)^{ - 1/m}}$ \cite{thornburg2016performance}.

\textit{Theorem 1.}\label{th1} 
	The successful transmission probability for typical D2D receiver  in millimeter-wave in $i$-th band is given by
	\begin{equation}\label{eq:12}
	\begin{array}{l}
	\mathbb{P} (SIN{R_{d,i}} \ge {T_{d,i}})= \sum\limits_{n = 1}^m 
	a_n\prod\limits_j {\exp \big( { - 2\pi {\lambda _{d,i}}A_j^D} \big)} \prod\limits_j {\exp \big( { - 2\pi {\lambda _{c,i}}A_j^C} \big)} {f_L}\big({R_{d,00,i}}\big)\\
	\,\,\,\,\,\,\,\,\,\,\,\,\,\,\,\,\,\,\,\,\,\,\,\,\,\,\,\,\,\,\,\,\,\,\,\,\,\,\,\,\,\,\,\,\,\,\,\,\,+ \sum\limits_{n = 1}^m b_n \prod\limits_j {\exp \big( { - 2\pi {\lambda _{d,i}}B_j^D} \big)} \prod\limits_j {\exp \big( { - 2\pi {\lambda _{c,i}}B_j^C} \big)} {f_N}\big({R_{d,00,i}}\!\big),
	\end{array}
	\end{equation}
	where $j \in \left\{ {LOS,NLOS} \right\}$, $ \lambda _{c,i} $ and $ \lambda _{d,i}$ represent the density of cellular users and D2D users in the $ i$-th band, respectively. Considering $k \in \left\{ {GG,Gg_s,g_sg_s} \right\}$, we have
	\begin{equation}\label{eq:13}
	\begin{array}{l}
	a_n={m \choose n} (-1)^{n+1}
	\exp \big( { - \frac{{an{T_{d,i}}R_{d,00,i}^{{\alpha _{\scaleto{L\mathstrut}{4pt}}}}}}{{{P_{d,i}}{G_{d,00}}}}{N_0}} \big)\\
	b_n={m \choose n} (-1)^{n+1} \exp \big( { - \frac{{an{T_{d,i}}R_{d,00,i}^{{\alpha _{\scaleto{N\mathstrut}{4pt}}}}}}{{{P_{d,i}}{G_{d,00}}}}{N_0}} \big)\\
	A_j^D = \sum\limits_k {{p_{k}}\int\limits_0^\infty  {\Big( {1 - {{{{\big( {1 + an\frac{{{T_{d,i}}R_{d,00,i}^{{\alpha _{\scaleto{L\mathstrut}{4pt}}}}{G_k}}}{{{r^{{\alpha _{\scaleto{L\mathstrut}{4pt}}}}}m{G_{d,00}}}}} \big)}^{-m}}}}} \Big){f_j}(r)rdr} }\\
	A_j^C = \sum\limits_k {{p_{k}}\int\limits_0^\infty  {\Big( {1 - {{{{\big(1 + an\frac{{{T_{d,i}}R_{d,00,i}^{{\alpha _{\scaleto{L\mathstrut}{4pt}}}}{P_{c,i}}{G_k}}}{{{P_{d,i}}{r^{{\alpha _{\scaleto{L\mathstrut}{4pt}}}}}m{G_{d,00}}}}\big)}^{-m}}}}} \Big){f_j}(r)rdr} }\\
	B_j^D = \sum\limits_k {{p_{k}}\int\limits_0^\infty  {\Big( {1 - {{{{\big( {1 + an\frac{{{T_{d,i}}R_{d,00,i}^{{\alpha _{\scaleto{N\mathstrut}{4pt}}}}{G_k}}}{{{r^{{\alpha _{\scaleto{N\mathstrut}{4pt}}}}}m{G_{d,00}}}}} \big)}^{-m}}}}} \Big){f_j}(r)rdr} } \\
	B_j^C = \sum\limits_k {{p_{k}}\int\limits_0^\infty  {\Big( {1 - {{{{\big( {1 + an\frac{{{T_{d,i}}R_{d,00,i}^{{\alpha _{\scaleto{N\mathstrut}{4pt}}}}{P_{c,i}}{G_k}}}{{{P_{d,i}}{r^{{\alpha _{\scaleto{N\mathstrut}{4pt}}}}}m{G_{d,00}}}}} \big)}^{-m}}}}} \Big){f_j}(r)rdr} }. 
	\end{array}
	\end{equation}
	$A_j^D$ and $A_j^C$ correspond to the LOS and NLOS interferences from D2D and cellular users, respectively when the desired signal is LOS, and  $B_j^D$ and $B_j^C$ correspond to the LOS and NLOS interferences from D2D and cellular users, respectively when the desired signal is NLOS.

\begin{proof}
	See Appendix. \hfill
\end{proof}

\textit{Theorem 2.} \label{th2}
	The successful transmission probability for a typical BS  in millimeter-wave in $i$-th band is as follows.
	\begin{equation}\label{eq:14}
	\begin{array}{l}
	{\mathop{\rm \mathbb{P}}\nolimits} (SIN{R_{c,i}} \ge {T_{c,i}})= \sum\limits_{n = 1}^m c_n \prod\limits_j {\exp \big( { - 2\pi {\lambda _{d,i}}E_j^D} \big)} \prod\limits_j {\exp \big( { -2\pi {\lambda _{c,i}}E_j^C} \big)} {f_L}\big({R_{c,00,i}}\big)\\
	\,\,\,\,\,\,\,\,\,\,\,\,\,\,\,\,\,\,\,\,\,\,\,\,\,\,\,\,\,\,\,\,\,\,\,\,\,\,\,\,\,\,\,\,\,\,\,\,+ \sum\limits_{n = 1}^m d_n\prod\limits_j {\exp \big( { - 2\pi {\lambda _{d,i}}F_j^D} \big)} \prod\limits_j {\exp \big( { - 2\pi {\lambda _{c,i}}F_j^C} \big)} {f_N}\big({R_{c,00,i}}\big),
	\end{array}
	\end{equation}
	where
	\begin{equation}\label{eq:15}
	\begin{array}{l}
	c_n={m \choose n} (-1)^{n+1} \exp \big( { - \frac{{an{T_{c,i}}R_{c,00,i}^{{\alpha _{\scaleto{L\mathstrut}{4pt}}}}}}{{{P_{c,i}}{G_{c,00}}}}{N_0}} \big)\\
	d_n={m \choose n} (-1)^{n+1} \exp \big( { - \frac{{an{T_{c,i}}R_{c,00,i}^{{\alpha _{\scaleto{N\mathstrut}{4pt}}}}}}{{{P_{c,i}}{G_{c,00}}}}{N_0}} \big)\\
	E_j^C = \sum\limits_k {{p_k}\int\limits_0^\infty  {\Big( {1 - {{{{\big( {1 + an\frac{{{T_{c,i}}R_{c,00,i}^{{\alpha _{\scaleto{L\mathstrut}{4pt}}}}{G_k}}}{{{r^{{\alpha _{\scaleto{L\mathstrut}{4pt}}}}}m{G_{c,00}}}}} \big)}^{-m}}}}} \Big){f_j}(r)rdr} } \\
	E_j^D = \sum\limits_k {{p_{k}}\int\limits_0^\infty {\Big( {1 - {{{{\big( {1 + an\frac{{{T_{c,i}}R_{c,00,i}^{{\alpha _{\scaleto{L\mathstrut}{4pt}}}}{P_{d,i}}{G_k}}}{{{P_{c,i}}{r^{{\alpha _{\scaleto{L\mathstrut}{4pt}}}}}m{G_{c,00}}}}} \big)}^{-m}}}}} \Big){f_j}(r)rdr} }\\
	F_j^C = \sum\limits_k {{p_{k}}\int\limits_0^\infty  {\Big( {1 - {{{{\big( {1 + an\frac{{{T_{c,i}}R_{c,00,i}^{{\alpha _{\scaleto{N\mathstrut}{4pt}}}}{G_k}}}{{{r^{{\alpha _{\scaleto{N\mathstrut}{4pt}}}}}m{G_{c,00}}}}} \big)}^{-m}}}}} \Big){f_j}(r)rdr} } \\
	F_j^D = \sum\limits_k {{p_k}\int\limits_0^\infty  {\Big( {1 - {{{{\big( {1 + an\frac{{{T_{c,i}}R_{c,00,i}^{{\alpha _{\scaleto{N\mathstrut}{4pt}}}}{P_{d,i}}{G_k}}}{{{P_{c,i}}{r^{{\alpha _{\scaleto{N\mathstrut}{4pt}}}}}m{G_{c,00}}}}} \big)}^{-m}}}}} \Big){f_j}(r)rdr} }  
	\end{array}
	\end{equation}

\begin{proof}
	The proof is similar to Theorem\ref{th1}. \hfill 
\end{proof}

By substituting STP expressions (\ref{eq:12}) and (\ref{eq:14}) into (\ref{eq:10}), the average rate for both D2D and cellular users  in millimeter-wave in $i$-th band is calculated. Now, we can compute the ASR of typical D2D receiver in millimeter-wave. 
\begin{equation}\label{eq:16}
AS{R_{d,i}} = {\lambda _{d,i}}{\bar R_{d,i}}.
\end{equation}

The EE of D2D users in multiple bands is defined as 
\begin{equation}\label{eq:17}
E{E_d} = \frac{{\sum\limits_i {AS{R_{d,i}}} }}{{\sum\limits_i {{\lambda _{d,i}}({P_{d,i}} + 2P_{cir}} )}},
\end{equation}
where $2P_{cir}$ is the circuit power of  both the D2D transmitter and receiver. 

We want to maximize the total EE by finding the optimum power of typical D2D transmitter in each band. We do not optimize the power of cellular users. So, our objective function and its constraints can be formulated as follows.
\begin{equation}\label{eq:18}
\begin{array}{l}
\mathop {\max }\limits_{{P_{d,i}}} \begin{array}{*{20}{c}}
{}&{E{E_d}}
\end{array} \\
s \cdot t \cdot \,\,\,\,\,\,\,\,\,(1)\,\,\sum\limits_{i = 1}^M {{P_{d,i}} \le {P_d}} \\
\begin{array}{*{20}{c}}
{}&&&&{(2)\,\,0 \le {P_{d,i}} \le {P_{d,i,\max }}},\,\forall i \in \left\{ {1,...,M} \right\}
\end{array}\\
\begin{array}{*{20}{c}}
{}&&&&{(3)\,\,\mathbb{P}(SIN{R_{d,i}} \ge {T_{d,i}})}
\end{array} \ge {\theta _{d,i}},\,\forall i \in \left\{ {1,...,M} \right\}\\
\begin{array}{*{20}{c}}
{}&&&&{(4)\,\,\mathbb{P}(SIN{R_{c,i}} \ge {T_{c,i}})}
\end{array} \ge {\theta _{c,i}},\,\forall i \in \left\{ {1,...,M} \right\}.
\end{array}
\end{equation}
As mentioned in Section II, the overall D2D users' transmission power should be less than the predefined threshold which is represented in the first constraint. The second constraint shows that the transmission power of devices in each band should not exceed the upper bound.  Constraint (3) and (4) are for satisfying the quality of service (QoS) of D2D and cellular users, respectively. It means that the STP of cellular and D2D users in $i$-th band should be higher than specific thresholds, which is represented by ${\theta _{c,i}}$ and ${\theta _{d,i}}$, respectively. In other words, when these constraints do not satisfy, the outage occurs. Due to both the objective function and QoS constraints, the problem is non-convex.
The integrals in (\ref{eq:13}) and (\ref{eq:15}) do not have closed-form solutions. Generally, we cannot derive closed-form formulas for the energy efficiency of D2D users and successful transmission probability. In other words, the transmission power of D2D users has no analytical solution. So, since it is analytically intractable, we solve the problem numerically. 

To solve the problem, MATLAB toolbox is used and non-linear programming function \textit{fmincon} is deployed \cite{MATLAB}. The interior point algorithm which is a numerical solver is used by fmincon. The solution is computed in a centralized manner.  We assume fixed transmission power for cellular users and consider that BS knows the cellular users' transmission power. The BS computes transmission power of a typical D2D user on each band.  The major steps of numerical power allocation algorithm is reviewed in Algorithm \ref{alg:1}. First, we initialize the fixed-value  parameters, such as cellular transmission power. Then, (\ref{eq:18}) is solved by \textit{fmincon} to find the optimal power of D2D users in $i$-th band. Finally, $EE_d$ is computed by substituting the optimal power of D2D users in (\ref{eq:17}).  

\begin{algorithm}[ht] 
	\caption{  Numerical power allocation algorithm \label{alg:1}}
	\begin{algorithmic}[1]  
		\State   Initialization: \mbox{\boldmath$P_c$},  \mbox{\boldmath$\lambda_d$}, \mbox{\boldmath$\lambda_c$}, \mbox{\boldmath$R_c$}, \mbox{\boldmath$R_d$}, \mbox{\boldmath${\theta _{d}}$}, \mbox{\boldmath${\theta _{c}}$}, $\beta$, $G$, $g_s$
		\State 		Solve (\ref{eq:18}) with \textit{fmincon} for given \mbox{\boldmath$P_c$} and obtain \mbox{\boldmath$P_{d}$} 
		\State   Compute $EE_d$ by substituting  \mbox{\boldmath$P_{d}$} in (\ref{eq:17})
	\end{algorithmic}
\end{algorithm}

For simplicity, the effect of thermal noise is neglected similar to other studies, i.e., \cite{9}, \cite{15}. In other words, we investigate the results in the interference limited regime. 
\section{Numerical results} \label{sec4}
In this section, we study the EE performance by using the analytical results which are obtained in the previous section.  Also, to verify our results, the Monte Carlo simulation is used. The  effective antenna gain for D2D and cellular users are assumed to be the same. So, ${G_{d,00}} = {G_{c,00}} = {G_0}$. The main system parameters which are utilized are in Table I. For validating our analytical results; total energy efficiency is obtained by averaging over 1000 realizations of the channel through Monte Carlo simulation. For simulations, we generate users with a Poisson distribution over an area of $3\,k{m^2}$. 
In addition, we consider different $m$'s (Nakagami parameter).  First, we assume $m=1$. In this case, the Nakagami distribution is converted to Rayleigh distribution.  By neglecting noise, $a_1=b_1=1$. By substituting $m$, $\alpha _N = 4$ and $\alpha _L = 2$ in (\ref{eq:13})  we have,
\begin{eqnarray}
\begin{array}{l}
A_j^D = \sum\limits_k {{p_k}\int\limits_0^\infty  {\Big(1 - {{\big(1 + \frac{{{T_{d,i}}R_{d,00,i}^2{G_k}}}{{{r^2}{G_{d,00}}}}\big)}^{ - 1}}\Big){f_j}(r)rdr} }  = \infty ,\\
A_j^C = \sum\limits_k {{p_k}\int\limits_0^\infty  {\Big(1 - {{\big(1 + \frac{{{T_{d,i}}R_{d,00,i}^2{P_{c,i}}{G_k}}}{{{P_{d,i}}{r^2}{G_{d,00}}}}\big)}^{ - 1}}\Big){f_j}(r)rdr} }  = \infty ,\\
B_j^D = \sum\limits_k {{p_k}\int\limits_0^\infty  {\Big(1 - {{\big(1 + \frac{{{T_{d,i}}R_{d,00,i}^4{G_k}}}{{{r^4}{G_{d,00}}}}\big)}^{ - 1}}\Big){f_j}(r)rdr} }  = \sum\limits_k {{p_k}\frac{\pi }{4}R_{d,00,i}^2\sqrt {\frac{{{T_{d,i}}{G_k}}}{{{G_{d,00}}}}} } \,,\\
B_j^C = \sum\limits_k {{p_k}\int\limits_0^\infty  {\Big(1 - {{\big(1 + \frac{{{T_{d,i}}R_{d,00,i}^4{P_{c,i}}{G_k}}}{{{P_{d,i}}{r^4}{G_{d,00}}}}\big)}^{ - 1}}\Big){f_j}(r)rdr} }  = \sum\limits_k {{p_k}\frac{\pi }{4}R_{d,00,i}^2\sqrt {\frac{{{T_{d,i}}{G_k}{P_{c,i}}}}{{{G_{d,00}}{P_{d,i}}}}} } \,.
\end{array}
\end{eqnarray}
By substituting these parameters in (\ref{eq:12}),  the successful transmission probability for typical D2D receiver in the $i$-th band is computed and (\ref{eq:17}) simplifies as follows.
\begin{equation}\label{eq:19}
E{E_d} = \frac{{\sum\limits_i {{s_i}\exp \Big( - 0.5{\pi ^2}R_{d,00,i}^2t\big({\lambda _{d,i}} + {\lambda _{c,i}}\sqrt {\frac{{{P_{c,i}}}}{{{P_{d,i}}}}} \big)\Big)} }}{{\sum\limits_i {{\lambda _{d,i}}\big({P_{d,i}} + 2{P_{cir}}\big)} }}
\end{equation}
where  $s_i\buildrel \Delta \over ={\lambda _{d,i}}{W_i}{{\log }_2}(1 + {T_{d,i}}){f_N}({R_{d,00,i}})$ and $t\buildrel \Delta \over = \sum\limits_k {{p_k}\sqrt {\frac{{{T_{d,i}}{G_k}}}{{{G_{0}}}}} } $.

Also, STP of the typical BS in $i$-th band simplifies to 
\begin{equation}\label{eq:20}
\begin{array}{l}
\mathbb{P}(SIN{R_{c,i}} \ge {T_{c,i}}) = \exp \Big( - 0.5{\pi ^2}R_{c,00,i}^2t\big( {{\lambda _{d,i}}\sqrt {\frac{{{P_{d,i}}}}{{{P_{c,i}}}}}  + {\lambda _{c,i}}} \big)\Big){f_N}\big({R_{c,00,i}}\big).
\end{array}
\end{equation}
Secondly, we assume $m = 2$. Similarly, a closed-form expression for EE of the D2D users  in millimeter-wave is obtained, by substituting $\alpha _N  = 4$ and $\alpha _L = 2$ in (\ref{eq:17}).
\begin{equation}\label{eq:21}
\begin{array}{l}
\frac{ASR_{d,i}}{{{\lambda _{d,i}}({P_{d,i}} + 2P_{cir})}}=\frac{{{\lambda _{d,i}}{W_i}{{\log }_2}(1 + {T_{d,i}})}}{{{\lambda _{d,i}}({P_{d,i}} + 2P_{cir})}}\times \\  
\!\Bigg[\!\!\bigg(\!\!2\!\exp \Big(\!\!\!- \!0.2102{\pi ^2}\!R_{d,00,i}^2t\big(\!{{\lambda _{d,i}}\! +\! {\lambda _{c,i}}\!\sqrt {\frac{{{P_{c,i}}}}{{{P_{d,i}}}}} } \big)\Big)\!\exp \Big(\! \!-\! 0.2974{\pi ^2}\!R_{d,00,i}^2t\big( {{\lambda _{d,i}} + {\lambda _{c,i}}\sqrt {\frac{{{P_{c,i}}}}{{{P_{d,i}}}}} } \big)\!\Big)\!\!\bigg)\!{f_N}({R_{d,00,i}}\!)\\
+ \bigg(2\exp \Big(\!\!\! - 0.707\pi R_{d,00,i}^4{t_1}\big( {{\lambda _{d,i}} + {\lambda _{c,i}}\frac{{{P_{c,i}}}}{{{P_{d,i}}}}} \big)\Big) \exp \Big(\!\!\! - 1.4142\pi R_{d,00,i}^4{t_1}\big( {{\lambda _{d,i}} + {\lambda _{c,i}}\frac{{{P_{c,i}}}}{{{P_{d,i}}}}} \big)\Big)\bigg){f_L}({R_{d,00,i}})\Bigg],
\end{array}
\end{equation}
where ${t_1} \buildrel \Delta \over = \sum\limits_k {{p_k}\frac{{{T_{d,i}}{G_k}}}{{{G_0}}}} $.

In this case, the STP of the typical BS  in millimeter-wave in $i$-th band is as follows.
\begin{eqnarray}\label{eq:22}
\begin{array}{l}
\mathbb{P}(SIN{R_{c,i}} \ge {T_{c,i}})= \\ 

\bigg(\!2\exp \Big(\!\!\! - 0.2102{\pi ^2}R_{c,00,i}^2t\big( {{\lambda _{c,i}} + {\lambda _{d,i}}\sqrt {\frac{{{P_{d,i}}}}{{{P_{c,i}}}}} } \big)\Big)\! \exp \Big(\!\!\! -\! 0.2974{\pi ^2}R_{c,00,i}^2t\big( {{\lambda _{c,i}} + {\lambda _{d,i}}\sqrt {\frac{{{P_{d,i}}}}{{{P_{c,i}}}}}} \big)\Big)\bigg){f_N}\big({R_{c,00,i}}\big)\\
+  \bigg(2\exp \Big(\!\!\! - 0.707\pi R_{c,00,i}^4{t_1}\big( {{\lambda _{c,i}} + {\lambda _{d,i}}\frac{{{P_{d,i}}}}{{{P_{c,i}}}}} \big)\Big) \exp \Big(\!\!\!-1.4142\pi R_{c,00,i}^4{t_1}\big( {{\lambda _{c,i}} + {\lambda _{d,i}}\frac{{{P_{d,i}}}}{{{P_{c,i}}}}} \big)\Big)\bigg){f_L}\big({R_{c,00,i}}\big).
\end{array}
\end{eqnarray}
Equations (\ref{eq:19}) and (\ref{eq:20}) are related to $m=1$ and are used for Rayleigh distribution. Equations (\ref{eq:21}) and (\ref{eq:22}) are associated with $m=2$ and are used for Nakagami distribution in our simulations.

\begin{figure}[ht]
	\centering
	\includegraphics[width=.5\linewidth]{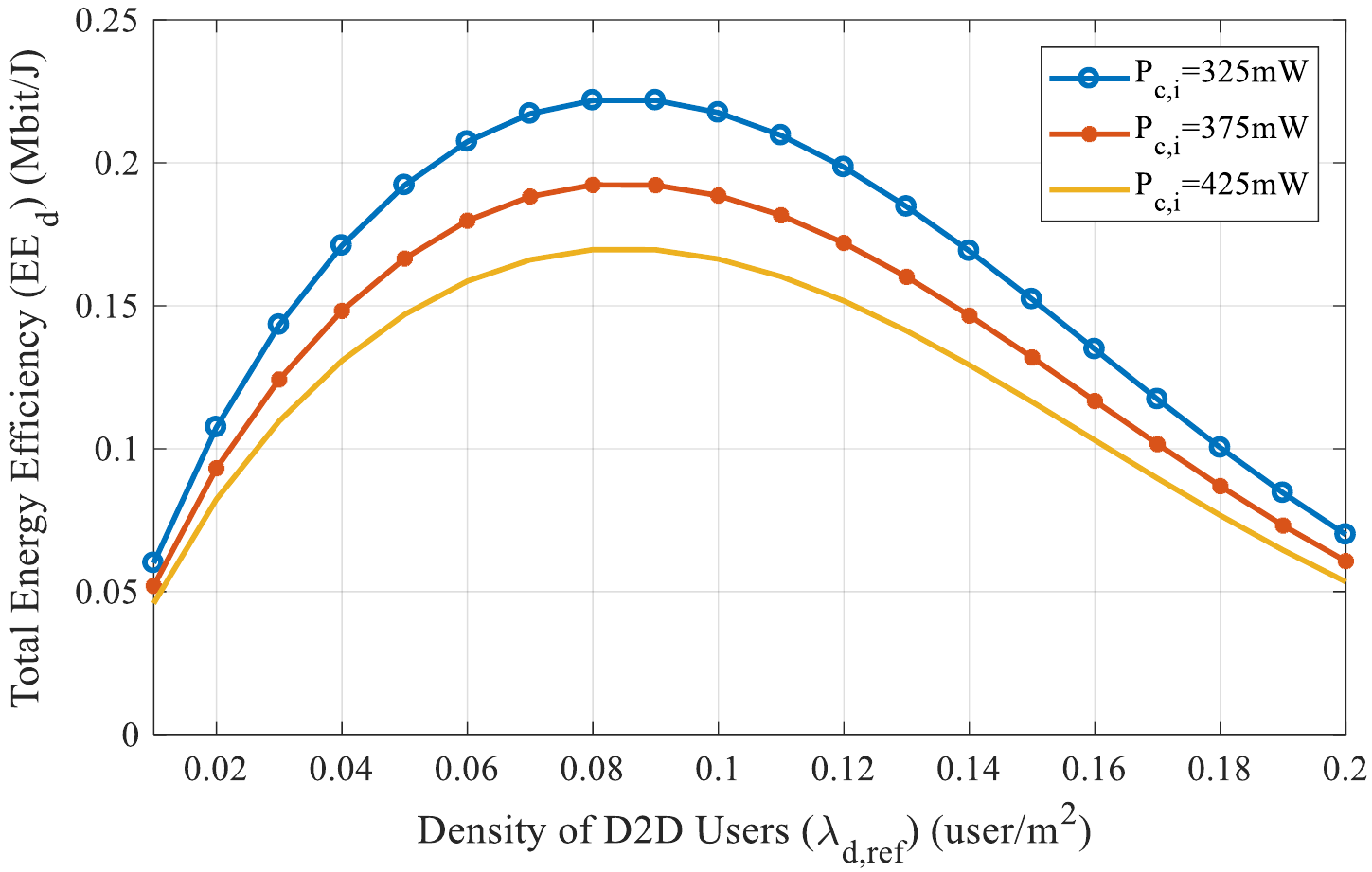}
	\caption{Total energy efficiency when outage occurs for $m=1$.}
	\label{fig. 2}
\end{figure}

\begin{figure}[ht]
	\centering
	\includegraphics[width=.5\linewidth]{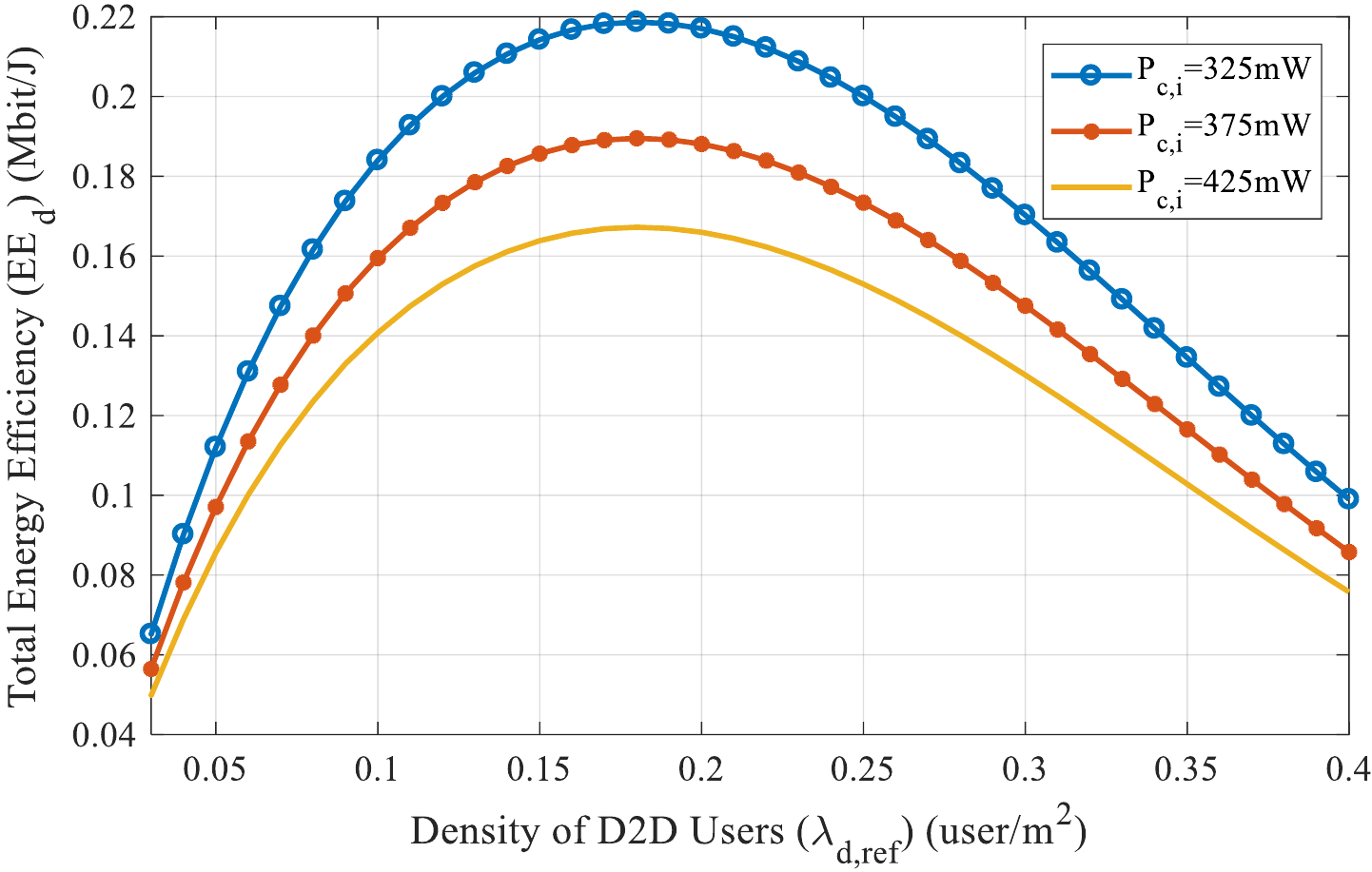}
	\caption{Total energy efficiency when outage occurs for $m=2$.}
	\label{fig. 3}
\end{figure}

Note that system parameters should be chosen carefully, otherwise outage occurs. For example, by increasing density of D2D users, the third and fourth constraints do not satisfy, so outage occurs. The results are shown in Fig. \ref{fig. 2} and Fig. \ref{fig. 3}. It is assumed that the channel is the best effort. It means that the channel does not provide any guarantees on final data rates.  As can be seen in these figures, by increasing the reference density of D2D users EE first increases and then decreases. This is because by increasing the density of D2D users their SIR rise too. However, in higher densities, the interference that cellular users produce is greater than the growth of ASR of D2D users. Therefore,  EE decreases. Also, by increasing $P_{c,i}$ the interference, by cellular users on the typical D2D receiver increases. Thus, the total energy efficiency decreases. 
\begin{table}[!t]
	\renewcommand{\arraystretch}{1.3}
	\caption{Simulation parameters}
	\label{table_example}
	\centering
	\begin{tabular}{lll}
		\hline\noalign{\smallskip}
		Parameter & Value  \\
		\noalign{\smallskip}\hline\noalign{\smallskip}
		$M$  & 5 \\
		$W_{i}$	& 20 MHz\\
		${\theta _{c,i}} $ & 0.95 \\
		${\theta _{d,i}} $  & 0.95 \\
		${T _{d,i}} $ & 0 dB \\
		${T _{c,i}} $ & 0 dB \\
		${R _{d,00,i}} $  & 10 m \\
		${R _{c,00,i}} $  & 30 m \\
		${\lambda _{d,i}} $ & ${10^{ - 4}}\,user/{m^2}$ \\
		${\lambda _{c,i}} $  & ${10^{ - 5}}\,user/{m^2}$ \\
		${P_{d,i,\max }} $ & 20 mW \\
		${P_{c,i}} $ & 325 mW \\
		${P_{cir}} $  & 0 mW \\
		${G} $& 10 dB \\
		${g_s} $  & 0.1 dB \\
		${\theta} $  & $\frac{\pi }{10}$  \\
		${\beta} $  & 0.45  \\
		\hline
	\end{tabular}
\end{table}
\begin{figure}[ht]
	\centering
	\includegraphics[width=.5\linewidth]{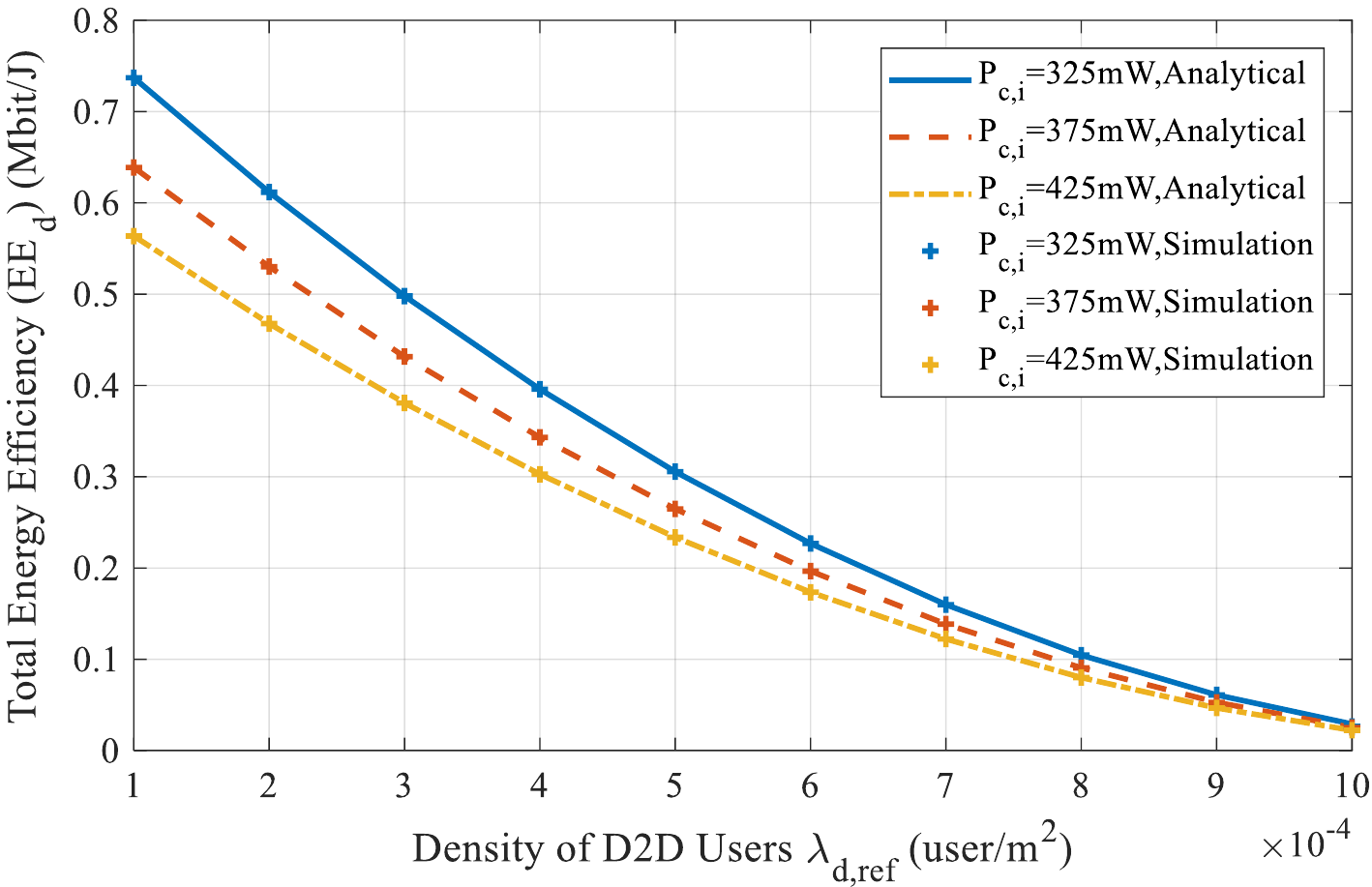}
	\caption{Total energy efficiency versus density of D2D users for Rayleigh distribution.}
	\label{fig. 4}
\end{figure}
\begin{figure}[ht]
	\centering
	\includegraphics[width=.5\linewidth]{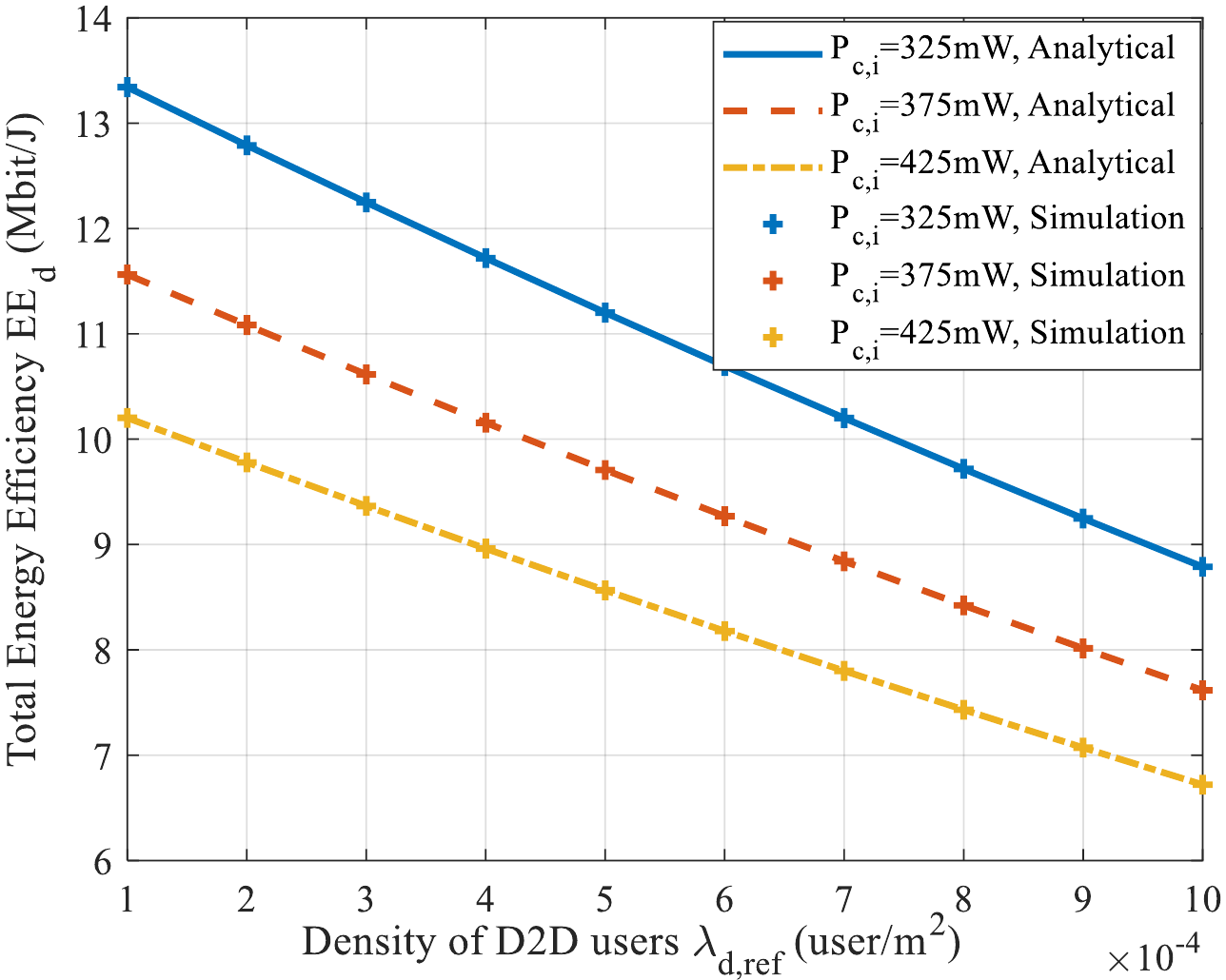}
	\caption{Total energy efficiency versus density of D2D users for Nakagami distribution with $m=2$.}
	\label{fig. 5}
\end{figure}

\begin{figure}[ht]
	\centering
	\includegraphics[width=.5\linewidth]{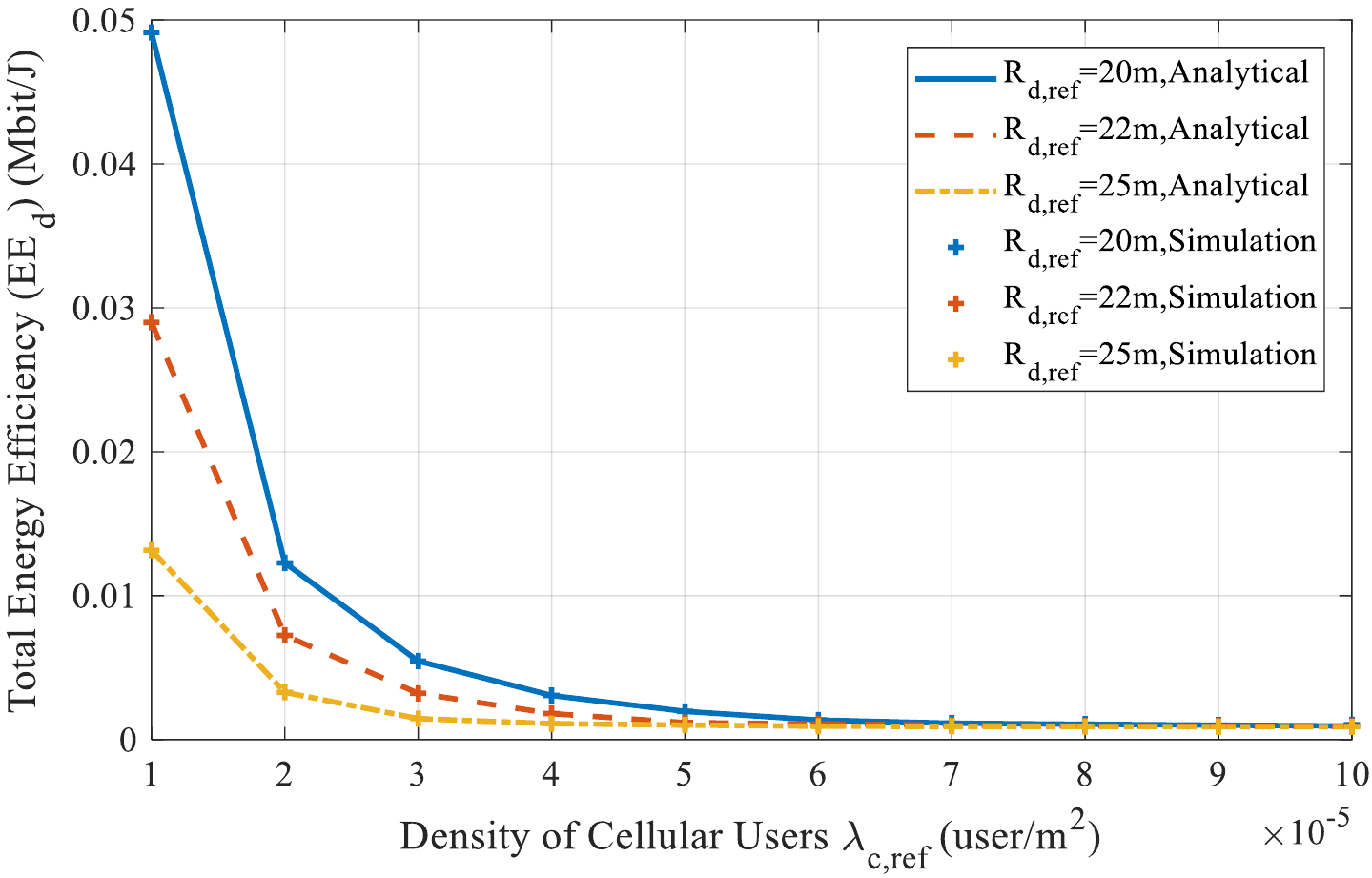}
	\caption{Total energy efficiency against density of cellular users for Rayleigh distribution.}
	\label{fig. 6}
\end{figure}
\begin{figure}[ht]
	\centering
	\includegraphics[width=.5\linewidth]{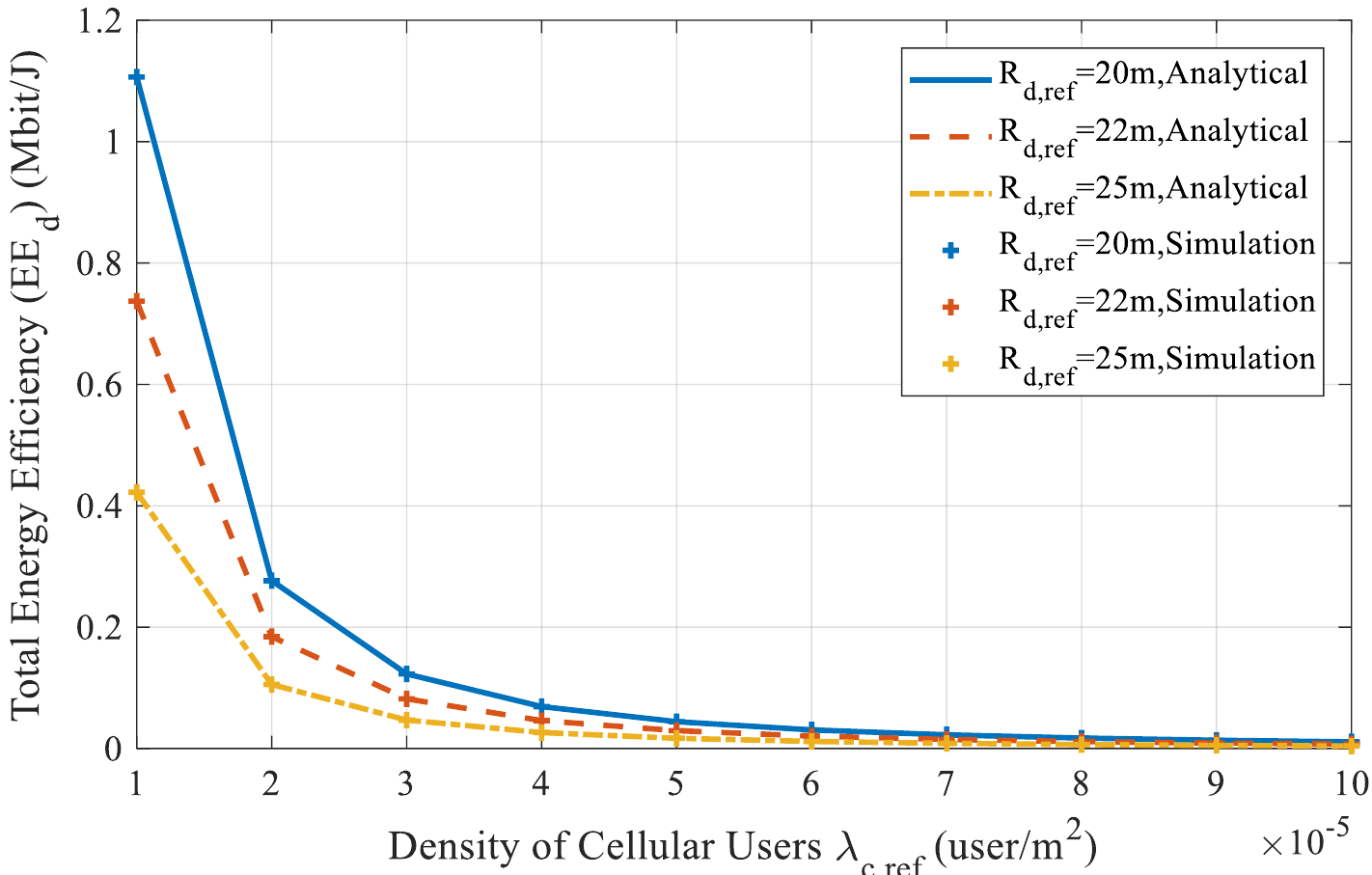}
	\caption{Total energy efficiency against density of cellular users for Nakagami distribution with $m=2$.}
	\label{fig. 7}
\end{figure}

Fig. \ref{fig. 4} and Fig. \ref{fig. 5} show total energy efficiency against the density of D2D users in two cases, $m=1$ and $m=2$. Here, the total power transmission of D2D users is 60 mW and ${\lambda _{d,i}} = [1,\,1,\,1,\,1,\,1] \times {\lambda _{d,ref}}$. The EE decreases by increasing the density of D2D users. This is because the growth of ASR of D2D users is less than the growth of interference that is produced by cellular users, by choosing these parameters. Similar to Fig. \ref{fig. 2} and Fig. \ref{fig. 3}, decreasing the transmission power of cellular users increases the total energy efficiency.

Comparing Fig. \ref{fig. 4} and Fig. \ref{fig. 5}, one can conclude that the total energy efficiency of channel gains with Nakagami distribution with $m=2$ is greater than that of it with Rayleigh distribution.  Since, the fading parameter $m$ ($m \ge 0.5$), which determines the shape of the distribution, varies as the fading condition ranges from severest ($m = 0.5$) to least ($m = \infty $),  Rayleigh fading ($m=1$)  causes more severe performance degradation than Nakagami fading with $m=2$.  When we have Nakagami distribution, we are working in millimeter-wave and the high-frequency regime, also, in the low-frequency regime, we are using Rayleigh distribution. Therefore, according to Fig. \ref{fig. 4} and Fig. \ref{fig. 5}, working in high frequencies is better than low frequencies. As can be seen in Fig. \ref{fig. 4} and Fig. \ref{fig. 5}, analytical and simulation results almost match.

We study the total energy efficiency performance under reference density of cellular users (${\lambda _{c,ref}}$) for $m=1$ and $m=2$. Three different distances for D2D users in all bands are set as 20 m, 22 m, and 25 m. The total power transmission of D2D users is 80 mW. Also, we consider ${\lambda _{c,i}} = [1,\,1,\,1,\,1,\,1] \times {\lambda _{c,ref}}$. The results are demonstrated in Fig. \ref{fig. 6} and Fig. \ref{fig. 7} which show that as the distance of D2D users increases, EE decreases, since the channel fading by the growth of distance becomes greater and the SIR decreases. Thus, the STP decreases, leading to a decrease in EE. 

Comparing Fig. \ref{fig. 6} and Fig. \ref{fig. 7}, again we can see that millimeter-wave outperforms operating in low frequencies. Another result is that by increasing ${\lambda _{c,ref}}$, EE decreases, since by increasing the number of cellular users the interference by cellular transmission increases. So, D2D users consume more power to coordinate this interference. Consequently,  EE decreases. Also, the simulation results match the analytical ones. 

\begin{figure}[t]
	\centering
	\includegraphics[width=.5\linewidth]{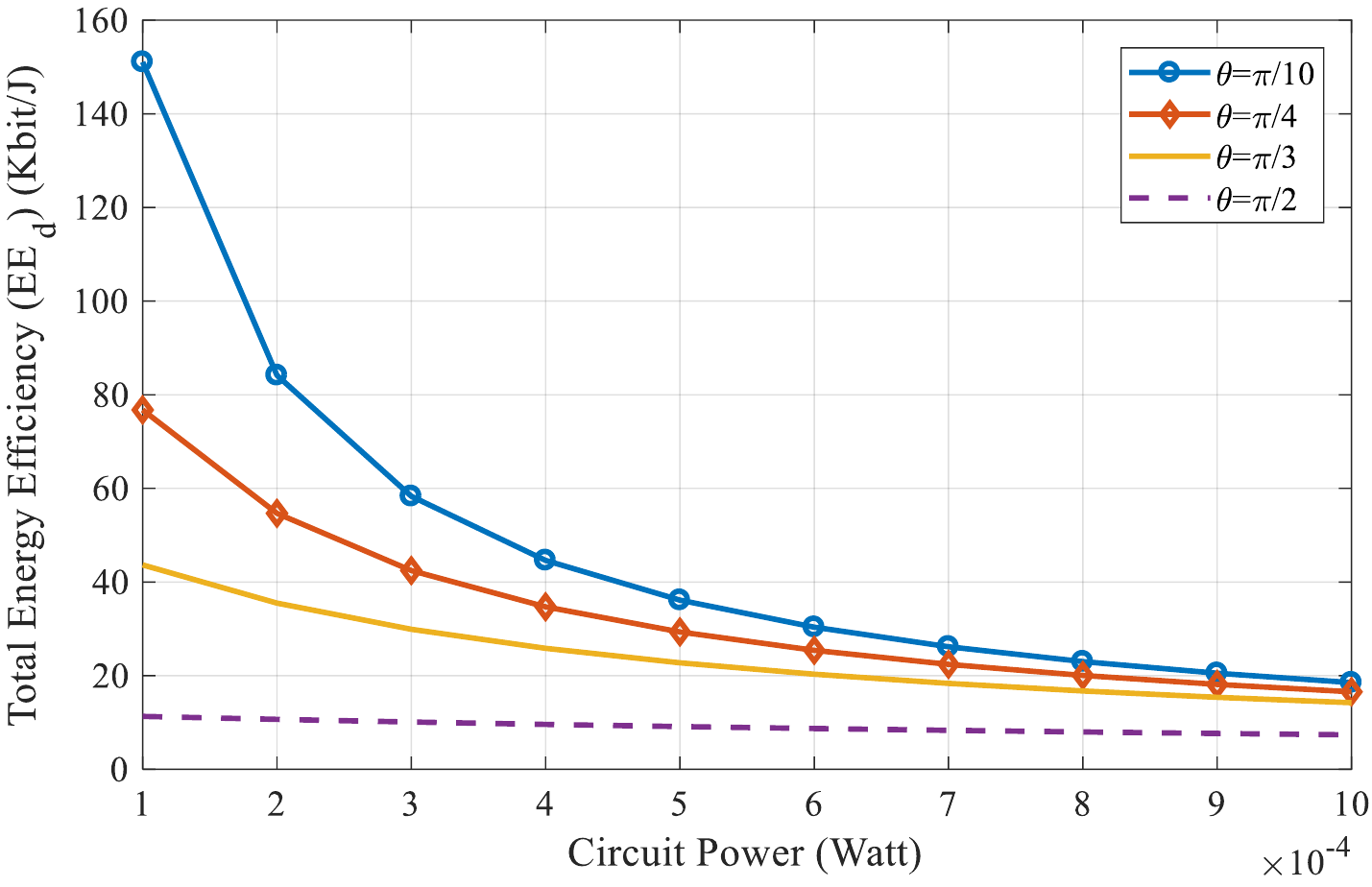}
	\caption{Total energy efficiency against circuit power for Rayleigh distribution.}
	\label{fig. 8}
\end{figure}

\begin{figure}[ht]
	\centering
	\includegraphics[width=.5\linewidth]{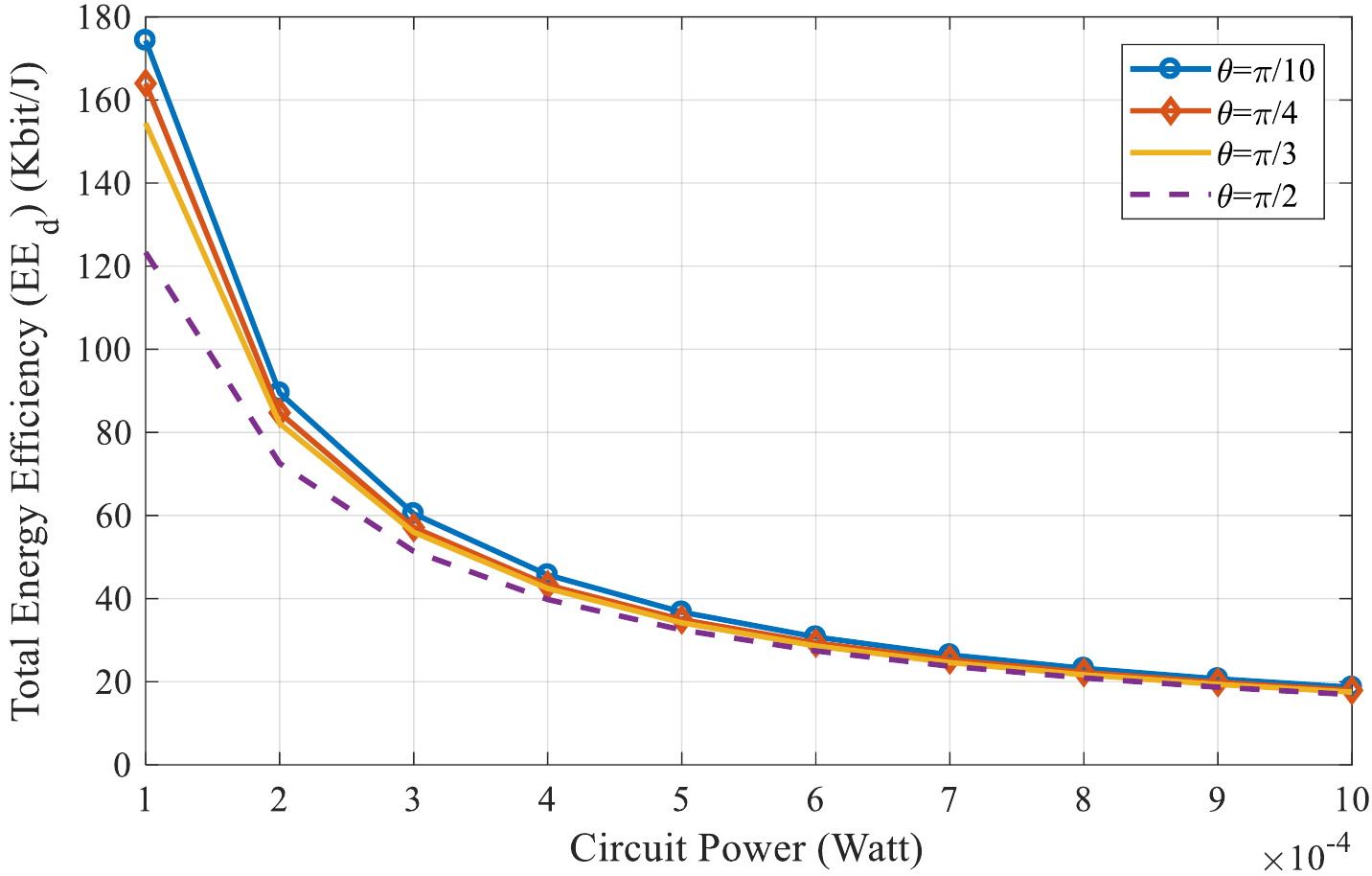}
	\caption{Total energy efficiency against circuit power for Nakagami distribution with $m=2$.}
	\label{fig. 9}
\end{figure}

The performance of energy efficiency is studied against circuit power in Fig. \ref{fig. 8} and Fig. \ref{fig. 9}. As the circuit power of D2D pairs is increased, the EE is decreased, since the power consumption of D2D users is increased and the circuit power is always a disadvantage in EE calculation. Also, the effect of the main lobe beamwidth on EE performance is investigated. As can be seen in Fig. \ref{fig. 8} and Fig. \ref{fig. 9}, by decreasing $\theta$, the main lobe has more power than side lobes, so  EE increases.

\begin{figure}[t]
	\centering
	\includegraphics[width=.5\linewidth]{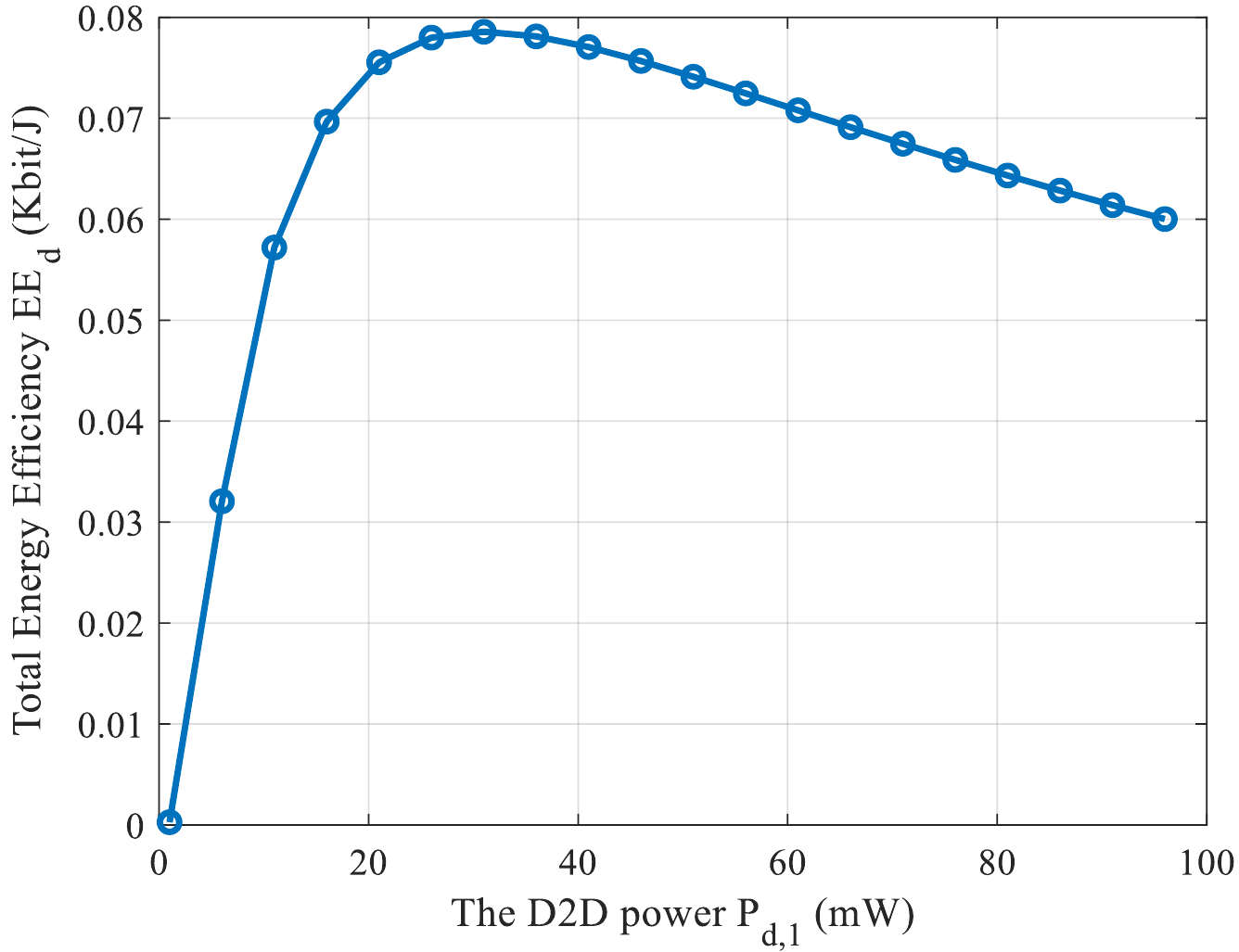}
	\caption{Total energy efficiency against $P_{d,1}$ for m=1.}
	\label{fig. 10}
\end{figure}

We simulate total energy efficiency against $P_d$ of the first band and show the result in Fig. \ref{fig. 10}. We assume $G=g_s=1 dB$ and $m=1$.
We can find that the EE rises at first and then declines as $P_{d,1}$ increases.
The reason  is that when $P_{d,1}$ is relatively small, the interference caused by spectrum sharing is slight. Thus, the growth of interference is insignificant, and EE increases. By increasing $P_{d,1}$, the interference which is produced by D2D users on cellular users becomes severe, so the EE decreases.

\section{Conclusions}
In this paper, the energy efficiency of D2D communication at mm-wave frequency band is maximized, considering the uniqueness of the mm-wave channel model such as directivity and blockage.  By considering the total EE of devices in the whole network as the objective function, the optimum power of the typical D2D user in each band is calculated. Also, stochastic geometry tools are used to obtain closed-form formulas for STP, ASR, and EE of D2D users in each band.  Network performance is investigated against reference  density of cellular and D2D users and circuit power. To evaluate the accuracy of these formulas, the Monte Carlo simulation is used. The channel fading is modeled by Nakagami distribution, and  the results are compared to Rayleigh distribution, which is a standard model for channel fading at low-frequency bands. From our results, one can conclude that mm-wave outperforms low-frequency bands. For future work, the total energy efficiency of the network can be maximized, and the optimum power of cellular users can be computed, as well. 
\section*{Appendix}

\textit{Proof of Theorem 1}
: Using the law of total probability, we have
\begin{equation}\label{eq:23}
\begin{array}{l}
\mathbb{P}(SIN{R_{d,i}} \ge {T_{d,i}}) ={\mathbb{P}_{L}}(SIN{R_{d,i}} \ge {T_{d,i}}){f_L}({R_{d,00}}) + {\mathbb{P} _N}(SIN{R_{d,i}} \ge {T_{d,i}}){f_N}({R_{d,00}}\!).
\end{array}
\end{equation}
Further, the STP for typical D2D receiver in the $i$-th band on the link being LOS can be evaluated as 
\begin{equation}\label{eq:24}
\begin{array}{l}
{{\mathop{\rm \mathbb{P}}\nolimits} _L}(SIN{R_{d,i}} \ge {T_{d,i}})
\\ = {{\mathop{\rm \mathbb{P}}\nolimits} _L}\Big( {{g_{d,00}} \ge \frac{{{T_{d,i}}R_{d,00,i}^{{\alpha _{\scaleto{L\mathstrut}{4pt}}}}}}{{{P_{d,i}}{G_{d,00}}}}\big( {{I_{d,c0,i}} + {I_{d,d0,i}} + {N_0}} \big)} \Big)\\= 1 - {\mathbb{P}_L}\Big({g_{d,00}} \le \frac{{{T_{d,i}}R_{d,00,i}^{{\alpha _{\scaleto{L\mathstrut}{4pt}}}}}}{{{P_{d,i}}{G_{d,00}}}}\big({I_\Phi } + {N_0}\big)\Big)\\ = 1 -\int\limits_0^\infty  {{\mathbb{P}_L}\Big({g_{d,00}} \le \frac{{{T_{d,i}}R_{d,00,i}^{{\alpha _{\scaleto{L\mathstrut}{4pt}}}}}}{{{P_{d,i}}{G_{d,00}}}}\big(x + {N_0}|{I_\Phi } = x\big)\Big)} {p_\Phi }\big(x\big)dx,
\end{array}
\end{equation}
where ${I_\phi }$  is the aggregate interference from cellular and D2D users to a typical D2D receiver in $i$-th band and ${p_\phi }$ is the probability distribution of PPP users. As mentioned in Section II, ${\phi_{c,i} }$ and ${\phi_{d,i} }$  are independent. Therefore, this probability is the product of probability distributions of D2D and cellular users. Also, we assume that each point in the building blockage is independent. So, the interference on typical D2D receiver by cellular and D2D users can be viewed separately as six independent PPPs such as 
\begin{equation}\label{eq:25}
\begin{array}{l}
{I_{d,c0,i}} = I_{d,c0,i,L}^{g_sg_s} + I_{d,c0,i,L}^{GG} + I_{d,c0,i,L}^{g_sG} + I_{d,c0,i,N}^{g_sg_s} + I_{d,c0,i,N}^{GG} + I_{d,c0,i,N}^{g_sG},\\
{I_{d,d0,i}} = I_{d,d0,i,L}^{g_sg_s} + I_{d,d0,i,L}^{GG} + I_{d,d0,i,L}^{g_sG} + I_{d,d0,i,N}^{g_sg_s} + I_{d,d0,i,N}^{GG} + I_{d,d0,i,N}^{g_sG}.
\end{array}
\end{equation}

We can approximate (\ref{eq:24}) as
\begin{equation}\label{eq:26}
\begin{array}{l}
{\rm \mathbb{P}_{L}}\big(SIN{R_{d,i}} \ge {T_{d,i}}\big)\\\mathop  \approx \limits^{(a)} 1 - {\mathbb{E}_\phi}\bigg[ {{{\Big( {1 - \exp \big( { - \frac{{a{T_{d,i}}R_{d,00,i}^{\alpha _{\scaleto{L\mathstrut}{4pt}}} }}{{{P_{d,i}}G_{d,00}}}X} \big)} \Big)}^m}}\bigg]\\\mathop  = \limits^{\left( b \right)} \sum\limits_{n = 1}^m {m \choose n} (-1)^{n+1} {\mathbb{E}_\phi}\Big[ {\exp \big( { - \frac{{an{T_{d,i}}R_{d,00,i}^{\alpha _{\scaleto{L\mathstrut}{4pt}}} }}{{{P_{d,i}G_{d,00}}}}X} \big)} \Big]\\\mathop  = \limits^{\left( c \right)} \sum\limits_{n = 1}^m {m \choose n} (-1)^{n+1} \exp \big({ - \frac{{an{T_{d,i}}R_{d,00,i}^{\alpha _{\scaleto{L\mathstrut}{4pt}}}}}{{{P_{d,i}G_{d,00}}}}{N_0}} \big) \prod\limits_j {\prod\limits_k {{\mathbb{E}_{{\phi _d}}}\Big( {\exp \big( { - \frac{{an{T_{d,i}}R_{d,00,i}^{{\alpha _{\scaleto{L\mathstrut}{4pt}}}}I_{d,d0,i}^{k,j}}}{{{P_{d,i}}{G_{d,00}}}}} \big)} \Big)}}\\ \,\,\,\,\,\,\,\,\,\,\,\,\,\,\,\prod\limits_j {\prod\limits_k {{\mathbb{E}_{{\phi _c}}}\Big( {\exp \big( { - \frac{{an{T_{d,i}}R_{d,00,i}^{{\alpha _{\scaleto{L\mathstrut}{4pt}}}}I_{d,c0,i}^{k,j}}}{{{P_{d,i}}{G_{d,00}}}}} \big)} \Big),} } 
\end{array}
\end{equation}
where $j \in \left\{ {LOS,NLOS} \right\}$, $k \in \left\{ {GG,Gg_s,g_sg_s} \right\}$, $X\buildrel \Delta \over = {{I_{d,c0,i}^{k,j}} + {I_{d,d0,i}^{k,j}} + {N_0}}$, (a) follows from Lemma 1, (b) is from binomial theorem and by considering that $m$ is  an integer and (c) is due to independent distributions of cellular and D2D users.

We  evaluate one of the expectation value in (\ref{eq:26}) as 
\begin{equation}\label{eq:27}
\begin{array}{l}
{\mathbb{E}_{{\phi_d}}}\Big( {\exp \big( { - \frac{{an{T_{d,i}}R_{d,00,i}^{\alpha _{\scaleto{L\mathstrut}{4pt}}} {I_{d,d0,i}^{GG,L}}}}{{{P_{d,i}G_{d,00}}}}} \big)} \Big)\\ = {\mathbb{E}_{{\phi_d}}}\Big( {\exp \big( { - \frac{{an{T_{d,i}}R_{d,00,i}^{\alpha _{\scaleto{L\mathstrut}{4pt}}} }}{{{P_{d,i}G_{d,00}}}}\sum\limits_{\ell  \in {\phi_{d,i}}} {{P_{d,i}}{g_{d,\ell 0}}GGR_{d,\ell 0,i}^{ - \alpha _{\scaleto{L\mathstrut}{4pt}}}} } \big)} \Big)  \\
={\mathbb{E}_{{\phi_d}}}\Big( {\prod\limits_{\ell  \in {\phi_{d,i}}} {\exp \big( { - an{T_{d,i}}R_{d,00,i}^{\alpha _{\scaleto{L\mathstrut}{4pt}} } {g_{d,\ell 0}}R_{d,\ell 0,i}^{ - \alpha _{\scaleto{L\mathstrut}{4pt}}}}{\frac{{GG}}{{{G_{d,00}}}}} \big)} } \Big)\\
\mathop  = \limits^{(a)} \exp \Bigg( {- 2\pi {\lambda _{d,i}}{{p_{GG}}}\int\limits_0^\infty  {\bigg( {1 - {\mathbb{E}_{{\phi_d}}}\Big( {\exp \big( Y \big)} \Big)} \bigg){{f_L}(r)}rdr} } \Bigg) \\
\mathop  = \limits^{(b)}\!\exp \bigg({ - 2\pi {\lambda _{d,i}}{{p_{GG}}}\int\limits_0^\infty  {\Big( {1 - {{{{\big( {1 + an\frac{{{T_{d,i}}R_{d,00,i}^{\alpha _{\scaleto{L\mathstrut}{4pt}}}}}{{{r^{\alpha _{\scaleto{L\mathstrut}{4pt}}}}m}}} \big)}^{-m}}}}} \Big){{f_L}(r)}rdr} }\bigg),
\end{array}
\end{equation}
where $Y\buildrel \Delta \over ={ - an{T_{d,i}}R_{d,00,i}^{\alpha _{\scaleto{L\mathstrut}{4pt}}} {g_{d,\ell 0}}{r^{ - \alpha _{\scaleto{L\mathstrut}{4pt}}}}}{\frac{{GG}}{{{G_{d,00}}}}}$, (a) is from the definition of Laplace functional and pgfl (probability generating functional) in stochastic geometry for Poisson point process  \cite{24}. In (\ref{eq:27}), (b) follows from the definition of moment generating function (MGF) ${M_x}(t) = \mathbb{E}\left( {{e^{ tx}}} \right)$. The MGF for  $\rm\Gamma (k,\theta )$ is ${(1 - t\theta )^{ - k}}$. Similarly, other expectation values can be computed and by substituting (\ref{eq:27}) into (\ref{eq:26}) the first summation in Theorem 1 is obtained. The same process obtains the second summation on the NLOS link. Then, these summations are multiplied by $f_L$ and $f_N$, respectively, and the proof ends.


\bibliographystyle{wileyNJD-AMA}%
\bibliography{ref}

\end{document}